# The Role of Smart Cities in Ethical Design Framework

Yijun Chen
Monash University, VIC, Australia

The integration of digital technologies into urban planning has given rise to "smart cities," aiming to enhance the quality of life and operational efficiency. However, the implementation of such technologies introduces ethical challenges, including data privacy, equity, inclusion, and transparency. This article employs the Beard & Longstaff framework to discuss these challenges through a combination of theoretical analysis and case studies. Focusing on principles of self-determination, fairness, accessibility, and purpose, the study examines governance models, stakeholder roles, and ethical dilemmas inherent in smart city initiatives. Recommendations include adopting regulatory sandboxes, fostering participatory governance, and bridging digital divides to ensure that smart cities align with societal values, promoting inclusivity and ethical urban development.

## 1.0 Introduction

Imagine a city where every piece of data collected enhances daily life while safeguarding individual privacy. This vision lies at the heart of smart cities—a transformative approach that integrates digital technologies with urban planning to improve economic efficiency, governance, and the quality of life for residents (Albino et al., 2015). Smart cities achieve this integration by deploying a variety of technologies across different sectors, fostering interconnected and efficient urban environments. For instance, Barcelona leverages the Internet of Things (IoT) for waste management, enabling real-time monitoring and optimization of waste collection processes (Maksimovic, 2020). Similarly, Singapore has developed integrated public transport systems that utilize smart traffic management technologies to streamline mobility and reduce congestion (Albino et al., 2015; Kitchin, 2016). These examples illustrate how smart technologies drive urban innovation and operational efficiency. Beyond infrastructure, smart cities employ advanced information systems (SIS) based on machine learning and artificial intelligence, coupled with big data analytics, to impact various aspects of personal and societal life (Kevin et al., 2019). A notable example is in healthcare: smart healthcare communities (SHCs) within smart cities utilize big data to transform the healthcare system by providing real-time access to information (Minyard, 2015). This enables more individuals to achieve greater treatment efficiencies and address healthcare time-lag issues, particularly during COVID-19 pandemic periods (Deloitte, 2022).

In addition to technological advancements, the pursuit of sustainable urban planning is a key objective of smart city implementation (Albino et al., 2015). Multiple literature indicates a notable shift toward human-centric, ethical, and sustainable approaches in smart city development. Foundational works such as Vazquez Brust's *"the City of Tomorrow"* and Musselwhite's study on human-centric urban design underscore the imperative for smart cities to prioritize residents' well-being and environmental sustainability (Musselwhite, 2022; Vazquez Brust, 2019).

However, a significant gap persists regarding how best to integrate ethically grounded governance principles into these technologically advanced urban systems (Caragliu et al., 2011; Townsend, 2013). This gap is particularly evident in discussions of data autonomy and security, equitable resource allocation, and the inclusivity of marginalized populations (Mark & Anya, 2019; Nam & Pardo, 2011a). To address this shortfall, our study applies the Beard and Longstaff (2018) framework which emphasizes self-determination, fairness, accessibility, and purpose—to critically assess contemporary smart city projects. Unlike traditional ethical theories such as utilitarianism (focused primarily on aggregate welfare utilitarianism (Sennett, 2018)) or deontological ethics (anchored to universal moral rules), Beard & Longstaff's multifaceted model integrates multiple dimensions simultaneously, providing novel insights into how diverse ethical imperatives can be operationalized in fast-evolving urban contexts (Beard & Longstaff, 2018).

By explicitly comparing how each principle in this framework addresses common blind spots in utilitarian or deontological approaches, this article demonstrates the framework's potential to foster more inclusive and transparent governance. In doing so, we aim to (1) evaluate the ethical trade-offs inherent in leading smart city implementations, (2) elucidate new findings on how advanced digital tools can both empower and marginalize different population segments, and (3) offer practical recommendations for policymakers and urban planners committed to ethically robust development. This study thereby positions itself at the intersection of emergent technology and ethical urban governance, highlighting the urgency of recalibrating "smartness" to encompass public well-being and fundamental rights.

## 2.0 Background

### *2.1 What is the "Smart City"*

The concept of smart cities has significantly evolved over the past decade, reflecting the multifaceted nature of urban innovation (Kitchin, 2016; Nam & Pardo, 2011b; Shelton et al., 2015a). Fundamentally, smart cities deploy information and communication technologies (ICT) to enhance residents' quality of life, improve operational efficiency, and promote sustainable development (Albino et al., 2015; Kitchin, 2016; WEF, 2021). Albino et al. (2015) define smart cities as urban areas that integrate ICT with urban planning to achieve economic efficiency, effective governance, and improved living standards. According to these authors, four key aspects characterize a "smart" city: (1) networked infrastructure that fosters political efficiency and supports social freedom and cultural development; (2) business-led urban development aimed at fostering innovation and economic growth; (3) social inclusion of diverse urban residents and the cultivation of social capital; and (4) environmental sustainability as a strategic consideration for future planning. This definition underscores the interdependent relationship between technology and urban infrastructure, emphasizing the pivotal role of digital innovation in addressing modern urban challenges.

Subsequently, Nam and Pardo (2011a) propose a more comprehensive framework that highlights the interplay between technology, people, and institutions in forming smart cities. Building on this, Vanolo (2014) critiques top-down approaches that overlook citizen empowerment, while Townsend (2013) underscores the importance of civic hacking and grassroots innovation, arguing that technological advancements should prioritize residents' welfare rather than corporate interests. Through this perspective, Kitchin (2016) adds that big data must serve explicit, community-centered objectives to avoid outcomes driven by profit motives or unchecked surveillance. In retrospect, Hollands (2008) called for reorienting smart city narratives around fairness and social equity, contending that digital initiatives should address, rather than deepen existing social disparities. Meijer & Bolívar (2016) and Shelton et al. (2015) reinforce this stance by illustrating how technology deployment can exacerbate or mitigate inequality, positioning fairness as a critical benchmark for genuinely "smart" urban development. Their findings also emphasize accessibility, highlighting that inclusive digital platforms and robust e-governance practices are vital for bridging digital divides. In support, Park et al. (2018) demonstrate how universal design principles and inclusive planning strategies can benefit residents of all backgrounds including older adults, individuals with disabilities, and other marginalized groups ensuring that new urban technologies effectively meet the diverse needs of a city's population. Collectively, these studies reveal that self-determination, purpose, fairness, and accessibility are not merely desirable add-ons but rather fundamental pillars for creating authentically smart cities.

Drawing inspiration from these foundational works, this article will focus on four critical elements of smart cities: self-determination, fairness, accessibility, and purpose. Each dimension corresponds to contemporary ethical challenges while aligning with the Beard & Longstaff framework. In the sections that follow, each principle is examined in its current application within smart cities and contrasted with its historical roots in more traditional urban settings. Through this comparative lens, we not only demonstrate how smart cities aim to address longstanding urban problems but also spotlight the novel ethical complexities that emerge as data and technology increasingly shape urban life.

### *2.2 Beard & Longstaff Framework*

In recent years, the governance of smart cities has become an increasingly complex endeavor, requiring ethical frameworks capable of addressing a wide spectrum of concerns arising at the intersection of technology, policy, and human rights (Caragliu et al., 2011; Meijer & Bolívar, 2016). Among the various models proposed to guide ethical deliberations in this space, the Beard & Longstaff framework has gained prominence for its holistic and integrative character (Beard & Longstaff, 2018). While many traditional ethical theories such as utilitarianism or deontological ethics offer valuable insights, they often prove insufficient when confronted with the multifaceted and dynamic challenges of contemporary urban environments (Sennett, 2018). Utilitarian approaches, for instance, tend to emphasize aggregate welfare and may inadvertently marginalize minority interests or ignore the subtlety of evolving privacy concerns within large-scale data-driven ecosystems (Richardson, 2024). Such an orientation can lead to "efficiency traps," where technological optimization is achieved at the expense of individual autonomy and vulnerable populations (Hollands, 2008). Similarly, deontological ethics, which prioritize adherence to universal moral rules, may overlook the contextual nuances of urban life. Rapid changes in technological capabilities and shifting policy landscapes mean that rigid rule-based systems can struggle to adapt, leading to ethical blind spots as new challenges emerge (Tseng & Wang, 2021).

In contrast, the Beard & Longstaff framework provides an adaptable, inclusive perspective that accommodates individual and collective values (Beard & Longstaff, 2018) Unlike theories that privilege a single dimension (e.g., outcomes or duties), it integrates personal autonomy, fairness, community welfare, and the overarching purpose of technological interventions. Moreover, the alignment of the Beard & Longstaff framework with "Ethical by Design" principles distinguishes it from more traditional models. Its emphasis on transparency, accountability, and non-discrimination directly resonates with global standards and best practices increasingly advocated by policymakers, civil society organizations, and industry consortia (Lifelo et al., 2024). These principles recognize that ensuring ethical governance in smart cities is not merely a matter of abstract moral reasoning; it requires actionable guidelines, stakeholder participation, and enforceable safeguards that protect user rights and enhance public trust (Beard & Longstaff, 2018). Such alignment ensures that the framework is not only theoretically sound but also practically applicable, enabling city planners, policymakers, and technologists to operationalize ethical principles in real-world scenarios (Meijer & Bolívar, 2016). Because of its flexibility and multidimensional focus, the Beard & Longstaff framework is especially effective for smart city governance. As global cities rapidly adopt digital infrastructures and data analytics, ethical frameworks must be dynamic enough to respond to new tools and shifting socioeconomic contexts. By providing a conceptual backbone and a normative guide, this framework helps city planners, policymakers, and other stakeholders uphold human values, foster equity, and sustain long-term viability in smart urban ecosystems.

## 3.0 Methods

This study used a conceptual and normative approach based on the Beard & Longstaff framework to identify the ethical challenges in smart city governance. The framework includes eight principles, but we only focused on four that are most relevant for smart city cases: self-determination, fairness, accessibility, and purpose. These principles were selected after reviewing academic literature, policy documents, and case studies, which highlighted these areas as critical to understanding the ethical dilemmas in smart cities, such as digital infrastructure, data governance, and stakeholder inclusion. As Figure 1 shown, we conducted a systematic literature review of peer-reviewed works retrieved from Google Scholar (n = 100) and Web of Science (n = 2635). After removing duplicates, we performed an initial relevance screening based on titles and abstracts, eliminating approximately 600 non-aligned studies. In the eligibility and quality assessment stage, 60 resources underwent scrutiny using the CRAAP (Currency, Relevance, Authority, Accuracy and Purpose) Test to gauge credibility, peer-review status to ensure scholarly rigor, language filters to maintain linguistic consistency, impact factor thresholds (greater than 1.0) to prioritize reputable journals, and elimination of publications lacking clear or evaluative methodologies.

**Figure 1:** Overview of research design

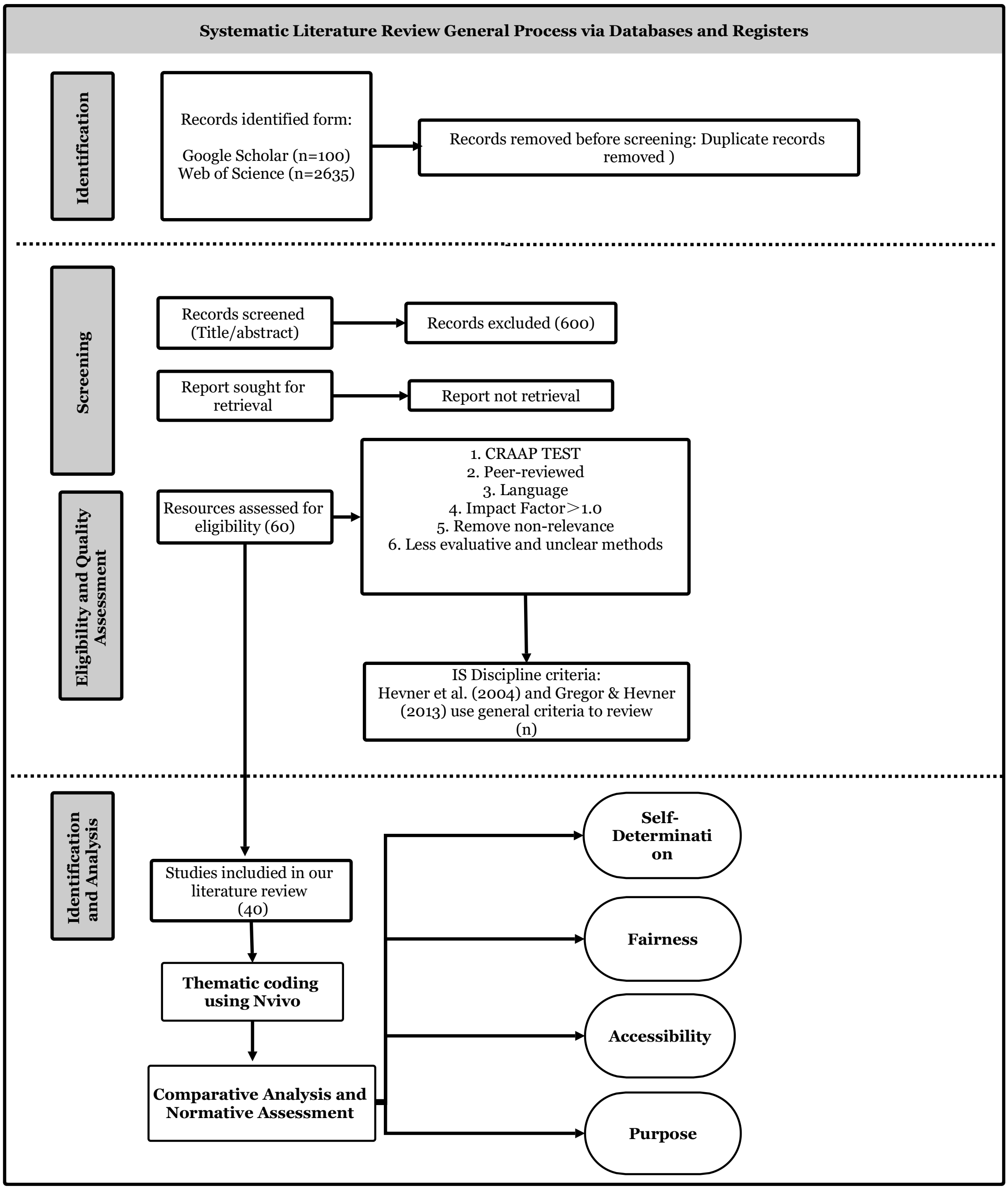

Following these assessments, we applied the IS discipline criteria from Gregor and Hevner (2013) as an additional layer of methodological rigor, concluding with a final corpus of 40 high-quality studies. We then employed NVivo for iterative reading and thematic coding, unveiling key ethical concerns like privacy, digital divides, unequal accessibility, and opaque governance. Mapping these issues against Beard & Longstaff's four principles illuminated how data governance and stakeholder involvement influence self-determination, fairness, accessibility, and purpose. Further comparative analysis of governance models, technological applications, and stakeholder dynamics spotlighted urgent ethical gaps—such as the absence of transparency, inequitable service distribution, and misalignment with public values. Taken together, these insights emphasize the critical need to strengthen ethical standards in emerging smart city contexts, providing actionable guidance for researchers, policymakers, and practitioners focused on embedding these four principles into future urban initiatives.

## 4.0 Ethical Challenges in Smart City Governance

### *4.1 Self-Determination*

According to Beard and Longstaff (2018), self-determination requires that individuals retain autonomy over their technological interactions. In many smart city settings, however, extensive data collection can undermine this autonomy and generate privacy risks. Although such systems aim to enhance efficiency and quality of life, they frequently compromise personal control over data. A recurring issue is the opacity surrounding data gathering, storage, and usage. For example, Singapore's smart traffic management system relies on continuous GPS data without offering clear opt-outs, prompting concerns about pervasive surveillance and potential abuse of sensitive location data (Degrande et al., 2021; Kitchin, 2013; Sennett, 2018). Similarly, in the United Kingdom, over 8,300 ANPR cameras capture approximately 30 million vehicle movements daily without informing the public or providing ways to limit their participation in data collection, thereby infringing on individuals' privacy rights (Degrande et al., 2021; Graham & Wood, 2003; Nam & Pardo, 2011a). By operating under the assumption that societal benefits such as improved safety and reduced congestion outweigh the ethical implications of diminished personal autonomy and privacy, these technologies subjugate individual rights to collective gains. Comparable dynamics are evident in New York City, where facial recognition-enabled CCTV cameras collect biometric data in public spaces without explicit consent (Brayne, 2021). Though often justified under security imperatives, this absence of agency regarding personal information has prompted public unease and criticism, underscoring the unchecked nature of these technological deployments and the potential for privacy breaches.Collectively, these examples illustrate how smart city initiatives frequently prioritize functional objectives over individual autonomy and privacy, leaving residents with minimal control over their personal data and interactions with urban systems. The academic literature corroborates these findings, highlighting the structural constraints embedded in many smart city designs (Keshavarzi et al., 2021; Meijer & Bolívar, 2016; Richardson, 2024; WEF, 2021). One example from Kitchin (2016) critiques the centralized, top-down governance approaches that often characterize these initiatives, arguing that they marginalize citizens and exclude them from meaningful decision-making regarding data use and privacy protections.

Similarly, Cath et al. (2017) contend that the prevailing focus on technological efficiency erodes self-determination and privacy, intensifying the power imbalances between individuals and the institutions that govern urban data flows. These studies emphasize that the erosion of self-determination and privacy is systemic rather than incidental, rooted in an inherent tension between centralized control and individual autonomy. In sum, the principles of self-determination and privacy are frequently compromised within the smart city paradigm. The pervasive, non-consensual data collection practices central to these systems highlight the ethical complexities at stake. As urban environments increasingly depend on real-time data-driven solutions, the tension between operational imperatives and respect for individual autonomy and privacy grows more acute. Without addressing these issues, the risk of eroding public trust and infringing on fundamental rights remains significant.

### *4.2 Fairness*

The principle of fairness, as outlined by Beard and Longstaff (2018), requires technological systems to treat all individuals equitably, regardless of socioeconomic background or geographic location. However, evidence drawn from case studies and comprehensive literature reviews reveals significant shortcomings in meeting this standard within smart city initiatives (Degrande et al., 2021; Graham & Wood, 2003; Heilweil, 2020; Sennett, 2018; WEF, 2021). One major challenge is that smart city technologies frequently favor middle- to upper-class residents who can readily access and utilize advanced digital tools (Cath et al., 2017; Mark & Anya, 2019; Nam & Pardo, 2011a). For instance, in New York City, smart kiosks intended to provide free internet access and information have been disproportionately placed in affluent neighborhoods, leaving lower-income communities with minimal coverage (Nam & Pardo, 2011a).

In contrast, smaller cities and rural regions encounter comparable challenges, as the resources and best practices successful in large metropolitan centers rarely translate effectively to less-developed or underserved locales (Cath et al., 2017; OECD, 2020). For example, in India, telemedicine services remain inaccessible to rural populations due to inadequate infrastructure and internet connectivity (eHealth, 2010). These scenarios highlight how certain groups are consistently left behind, undermining any claims of inclusivity inherent in smart city rhetoric. The systemic nature of these inequities is further highlighted by current AI perspectives. Kontokosta and Hong (2021) provide empirical evidence of significant bias in resident-reported data by examining socio-spatial disparities in '311' complaint behavior in Kansas City, Missouri.

Moreover, the deployment of automated systems, such as the “Robodebt” scheme implemented by the Australian government, exemplifies how technology can perpetuate systemic injustices when not carefully designed and monitored (Sheehy & Ng, 2024). The Robodebt system relied heavily on data matching and algorithms to identify and reclaim alleged overpayments of welfare benefits. However, it resulted in numerous wrongful debts being imposed on vulnerable and marginalized individuals, highlighting the potential for automated systems to exacerbate inequalities and undermine fairness.

Consequently, the principle of fairness is frequently compromised within smart city development. By catering predominantly to affluent stakeholders and neglecting economically or geographically disadvantaged communities, these initiatives risk deepening existing inequalities rather than alleviating them. Without deliberate strategies to ensure more equitable access and benefit distribution, smart cities may ultimately undermine their own goals of inclusivity, sustainability, and social justice

### *4.3 Accessibility*

The principle of accessibility, as defined by Beard and Longstaff (2018), emphasizes the ethical obligation for smart city technologies to be designed to guarantee equitable access for all individuals, including those with disabilities, the elderly, and residents in marginalized communities. Based on AHRC (Australian Human Right Commission) guidance, Farthing et al. (2019) further emphasize that new technologies should uphold the right to increased accessibility, ensuring that all vulnerable populations are not marginalized by digital advancements. However, we observed current empirical literature and case studies reveal substantial gaps in the current implementation of accessibility within smart city frameworks (Khromova et al., 2024; Mark & Anya, 2019; Sánchez-Corcuera et al., 2019; Tan & Taeihagh, 2020).

Our literature review indicates that the existing challenges of accessibility in both the public and private sectors remain underexplored, a finding consistent with the results of Lim et al. (2018) and Zakir et al. (2024).For example, Szewczenko (2020) explores how Information and Communication Technologies (ICT) can enhance urban accessibility and quality for the elderly people. However, there is no integrated framework to consider generalization across diverse urban contexts. On the other hand, Mora et al. (2016) present a mobility system that monitors urban accessibility for individuals, but collecting location data from users raises important privacy and security concerns. Another problem is lack of consideration for the institutional and legal framework as it pertains to actual cases. Compare to our case study, we observed a primary barrier to accessibility in smart cities is the geographic and economic disparity that hinders equitable service distribution (Sánchez-Corcuera et al., 2019; Tan & Taeihagh, 2020). Kitchin (2013) identifies these challenges are largely due to inadequate infrastructure and limited technical capabilities necessary for the physical deployment of advanced technologies. For instance, in Detroit, USA, approximately 29.71% of the population lacks reliable broadband access, effectively excluding them from the benefits of smart city services such as digital governance platforms, smart healthcare systems, and educational technologies (NDIA, 2018). This lack of infrastructure not only impedes access but also perpetuates existing socioeconomic inequalities, as marginalized communities remain disconnected from the digital advancements that could enhance their quality of life (García & Kim, 2020; Nam & Pardo, 2011b; Richardson, 2024). Multiple extant literature argued urban centers often results in a skewed distribution of technological resources, where affluent neighborhoods enjoy advanced smart infrastructures while economically disadvantaged areas are left behind (Button & Taylor, 2000; X. Ma et al., 2024; Richardson, 2024). As such, Woyke (2019) and Kolotouchkina et al. (2022) highlight that as smart city technologies become more sophisticated, the gap between privileged "smart citizens" and those marginalized by socioeconomic and geographic factors continues to expand. In India, telemedicine platforms introduced to provide healthcare services remotely were inaccessible to rural populations due to inadequate internet infrastructure and insufficient training on digital tools (Randell-Moon & Hynes, 2022; Tan & Taeihagh, 2020).

Also, the widening digital divide has been a focal point of concern among researchers and conference presentations. Woyke (2019) and Kolotouchkina et al. (2022) highlight that as smart city technologies become more sophisticated, the gap between privileged "smart citizens" and those marginalized by socioeconomic and geographic factors continues to expand. This disparity results in serious inaccessibility issues, where the benefits of smart city initiatives are disproportionately enjoyed by those who possess the means and knowledge to leverage advanced technologies. For example, in Nairobi, Kenya, smart transportation solutions such as app-based ride-hailing services are predominantly utilized by middle and upper-class residents, while low-income individuals reliant on traditional public transportation remain excluded due to cost barriers and lack of access to smartphones (Kolotouchkina et al., 2022). Furthermore, the COVID-19 pandemic has illuminated and exacerbated accessibility challenges within smart cities. During the pandemic, many cities deployed digital public services to facilitate contactless interactions and remote operations. However, these systems often lacked features to accommodate individuals with limited digital literacy or cognitive impairments. In India, telemedicine platforms introduced to provide healthcare services remotely were inaccessible to rural populations due to inadequate internet infrastructure and insufficient training on digital tools (Randell-Moon & Hynes, 2022; Tan & Taeihagh, 2020) Similarly, in Melbourne, Australia, the rapid implementation of smart contact tracing applications faced criticism for excluding elderly populations who were less familiar with smartphone technology, thereby limiting their ability to benefit from these public health measures (Randell-Moon & Hynes, 2022).

While the principle of accessibility is foundational to the ethical deployment of smart city technologies, significant gaps remain in ensuring that all populations can equally benefit from these advancements. Geographic and economic barriers, coupled with systemic oversights in design and implementation, continue to impede the realization of truly inclusive smart cities. The persistent digital divide, as evidenced by reviewed cases in Detroit, Nairobi, and Melbourne, underscores the urgent need for more equitable infrastructure distribution and user-centered design approaches in future smart city initiatives (S. Park, 2017). Addressing these challenges is essential not only for upholding ethical standards but also for fostering inclusive, sustainable, and resilient urban environments.

### *4.4 Purpose*

The principle of purpose, as established by Beard and Longstaff (2018), underscores the imperative for technological systems to be constructed with integrity, transparency, and clearly delineated goals. These principles require that smart city technologies possess transparent objectives that correspond with people' actual requirements for smart urban design. Our systematic literature review, which includes various case studies and theoretical frameworks, consistently reveals that smart cities frequently lack clear, universally accepted definitions, thus constraining their practical alignment with societal needs (Kolotouchkina et al., 2022; Nam & Pardo, 2011a; Sánchez-Corcuera et al., 2019). Current literature revealed various gaps between citizens' perception of the term "Smart" in the context of smart cities and their expectations for its implications. The disconnect is hindering the successful development of smart cities, preventing them from realizing their full potential (Cavada et al., 2014). This issue likely arises from stakeholders having ambiguous and divergent perceptions of what constitutes urban planning (de Gooyert et al., 2017). In the absence of a unified vision, misinterpretations may hinder the realization of the objective to enhance urban intelligence. In Copenhagen, the substantial initial expenses of renewable infrastructure, restricted scalability of specific technologies, and regulatory obstacles have impeded advancement (Negro et al., 2012). Likewise, cities such as Dubai and Songdo prioritize advanced infrastructure aimed at economic development and futuristic settings, yet frequently neglect to directly associate these technologies with the quality of life of their inhabitants (Chang, 2021; Keshavarzi et al., 2021; Nam & Pardo, 2011a). These instances demonstrate that a lack of a definitive objective yields substantial repercussions. Stakeholders, comprising residents and municipal officials, may misunderstanding the advantages and limitations of these initiatives (Taylor et al., 2020). When cities promote themselves as "smart" without providing concrete advantages, disengagement or disillusionment may ensue. This highlights the significance of a collective vision and framework.

However, when frameworks for elucidating such objectives are insufficient, the intricacy of governance becomes apparent. Unfortunately, The OECD statistics indicates that 42% of municipal administrative entities lack a standardized framework for assessing digital infrastructures (OECD, 2021). Additionally, approximately 30% of mid-sized cities globally lack a formal governance framework for coordinating smart technology initiatives (UN-Habitat, 2021). This institutional ambiguity frequently results in disjointed projects, misaligned funding, and sluggish policy adaptation, collectively intensifying what scholars refer to as the "complexity of governance" within smart city ecosystems. These statistics underscore both operational difficulties such as identifying areas for network expansion and sociopolitical obstacles, including the need to balance inclusivity, data privacy, and resource distribution.

## 5.0 Discussion and Implications

The findings from this study reveal a multifaceted landscape in the governance of smart city systems, encompassing a wide range of key issues identified include privacy, transparency, digital inclusion, and the roles of various stakeholders (Cath et al., 2017; OECD, 2021). These challenges highlight the complexity of governance in smart cities, where emerging technologies must be integrated with ethical and inclusive governance models. Before delving into a detailed discussion, Figure 1 offers a visual representation of the research gaps identified in this study, alongside corresponding recommendations to address these gaps. The diagram maps key challenges in smart city governance, including digital inclusion, privacy regulations, public-private partnerships (PPPs), and adaptive urban systems, as well as the emphasize on role of Freedom of Information (FOI) laws in enhancing transparency and accountability. This visualization clarifies the interconnectedness of various challenges and provides a comprehensive overview of how the proposed recommendations integrate within the broader governance framework for smart cities. By presenting a structured roadmap of these relationships, the diagram serves as a conceptual guide for future research, highlighting areas that require further exploration and the solutions proposed in this study. With this visual overview established, the discussion proceeds to examine the key findings, their implications, and how they relate to the research gaps and recommendations illustrated in the diagram.

**Figure 2** Recommendations for addressing our existing ethical challenges

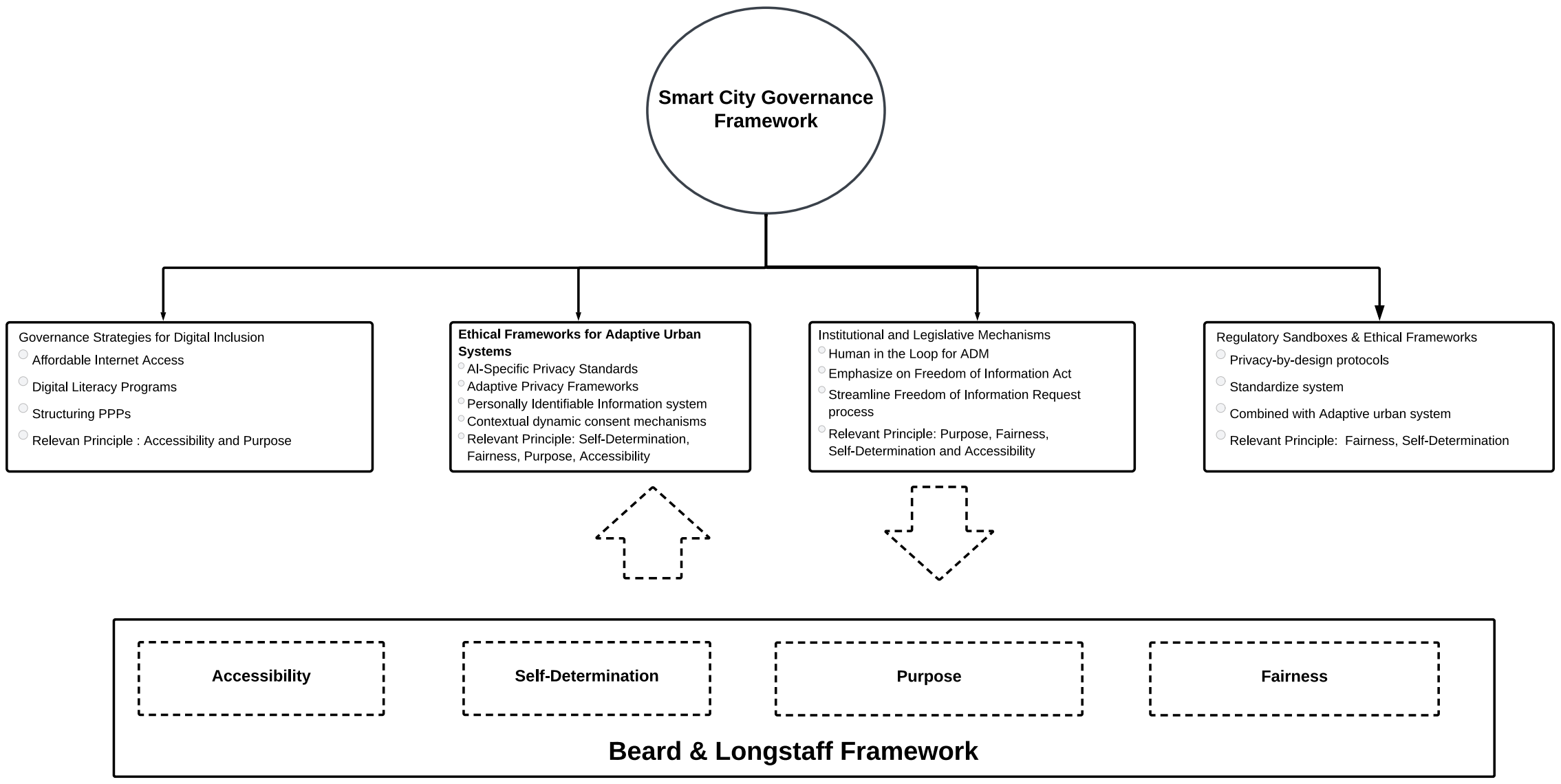

### *5.1 Governance Strategies for Digital Inclusion*

A core gap highlighted by this study is the absence of governance strategies that systematically prioritize digital inclusion. Although smart city technologies can improve urban management, public safety, and service delivery, they frequently intensify digital divides in underserved communities(Kamtam, 2023). In these areas, access to technology and digital literacy are frequently limited, hindering individuals from fully participating in the opportunities offered by smart cities (Kamtam, 2023; Tan & Taeihagh, 2020). To address this gap, our study emphasizes several strategic recommendations. First, expanding affordable internet access through public-private partnerships is essential to ensure that underserved communities have reliable and affordable high-speed internet. For example, international efforts like Google's Project Loon aims to facilitate connectivity through the deployment of balloons. In 2013, engineers launched 30 balloons over New Zealand to assess connectivity potential, providing antennas to residents to enable Internet access via these balloons(Serrano et al., 2023). In addition, digital literacy programs developed in collaboration with educational institutions and community organizations can equip residents with the skills necessary to effectively engage with smart city technologies. These programs foster greater participation and reduce the digital competency gap. For example, in India, instructional classes train adults (especially rural dwellers, senior citizens, and poor people) on internet usage. Through digital platforms like the National Knowledge Network, AADHAR, and eSeva, these groups can access critical information about business opportunities and market conditions, bringing up-to-date insights to their fingertips (Swargiary & Roy, 2023).

### *5.2 Ethical Frameworks for Adaptive Urban Systems*

Based on our findings above, we recommend adaptive urban systems are vital for addressing the complex and dynamic challenges of smart cities. These systems utilize real-time data and AI-driven decision-making to deliver flexible and responsive solutions to urban issues, aligning with the principles of inclusivity, sustainability, and accountability (Beard & Longstaff, 2018; Kitchin, 2016). Our research demonstrates that adaptive systems outperform static models in unpredictable urban environments by dynamically reallocating resources and optimizing operations, thereby enhancing efficiency and resilience (Mark & Anya, 2019). For instance, their ability to evolve with societal needs ensures long-term functionality and relevance, addressing the limitations of traditional urban governance frameworks (Nam & Pardo, 2011b). Crucially, the adoption of adaptive urban systems must embed ethical principles to ensure privacy, inclusivity, and transparency. Privacy safeguards such as anonymization and differential privacy are essential for protecting individual rights while enabling system functionality (Saini et al., 2011; Ziegeldorf et al., 2013).

In order to sustain adaptive urban system, we find Göbel & Kitzing (2023) introduce a new approach to improve the verification and communication of processing of Personally Identifiable Information ( PII ). In the context of a smart city, the Privacy-Enhancing Verification Component (PE-VC) is highly relevant because it ensures that all data-driven services remain compliant with privacy regulations while supporting a transparent, citizen-centric approach. By leveraging evidence from Germany, where data protection officers rely on the Compliance Assertion Language (COMPASS) to define and uphold legal anonymity requirements, PE-VC reduces the need for granular insights into the exact de-identification steps. This streamlines compliance verification without additional software intervention, thereby minimizing privacy risks and administrative overhead. As a result, city administrators can integrate a wide range of smart services—such as traffic management or public health analytics—knowing that personal data is securely protected and handled in accordance with established legal frameworks, ultimately fostering greater citizen trust and participation in smart city initiatives (Anthopoulos et al., 2023).

Another approach including contextual dynamic consent mechanisms offer a novel view to address freedom and privacy concerns while maintaining smart system adaptability. These AI-driven systems allow residents to adjust their privacy preferences in real time based on specific contexts, such as consenting to data sharing during public emergencies but opting out in routine scenarios (Herrera et al., 2022). This ensures that consent remains aligned with individual values and evolving societal conditions, fostering trust and engagement with adaptive technologies. To further enhance equity, our findings highlight the importance of equity-driven urban simulation models as a critical tool for implementing adaptive systems. These AI-powered simulations evaluate the societal impacts of proposed systems before deployment, modeling outcomes for diverse demographic groups to identify potential inequities (Ryerson et al., 2022). By simulating resource distribution, accessibility, and socio-economic impacts, these models ensure that our adaptive systems are designed to promote fairness and inclusivity across all segments of the population. Such tools align with the principles of participatory governance and allow policymakers to preemptively address disparities, ensuring that adaptive urban systems contribute to equitable and sustainable urban development.

### *5.3 Regulatory Sandboxes: A Mechanism for Testing and Refining Governance Models*

Another recommendation combined with adaptive urban system, we advocate Regulatory sandboxes can provide a novel, time-bound environment for testing and refining governance models and emerging technologies in smart cities, ensuring that new initiatives align with ethical considerations (as we identified above) while managing risk (Tan & Taeihagh, 2020). By allowing pilot deployments in controlled scopes and timelines, policymakers remain responsive to fast-evolving technology—for instance, testing drone logistics or autonomous vehicles—while collecting real-world data to assess performance, safety, and user acceptance (Raghunatha et al., 2023). Meanwhile, as we discussed before, Privacy-by-design protocols, such as anonymization and Privacy-Enhancing Technologies, can be fine-tuned without exposing the entire population to potential surveillance risks, exemplified by Singapore's geofenced sensor trials (Tan & Taeihagh, 2020). In parallel, citizen engagement—as demonstrated in Barcelona's participatory IoT sandbox—fosters trust and transparency, empowering community members to shape data-handling guidelines. This also extend to another case study including Helsinki's MaaS pilot project for traffic routing, benefit from iterative algorithmic refinements that address concerns like bias or uneven service allocation (Kamargianni et al., 2016). Limiting experimentation geographically or functionally, as seen in Rotterdam's "Smart Port," also lowers financial and operational risks, enabling decision-makers to analyze cost-benefit outcomes before committing to large-scale adoption(Puussaar et al., 2018). In terms of policy, we advocate for the standardization of regulatory sandboxes to ensure that emerging technologies comply with privacy, ethical, and governance standards, thereby fostering more resilient, citizen-centric, and transparent smart city ecosystems. In the United Kingdom, the Financial Conduct Authority (FCA) established standardize regulatory sandbox to evaluate AI-driven financial advisory tools (Truby, 2020). The AI applications were assessed for biases and fairness, guaranteeing that they delivered equitable financial advice to all users irrespective of their socioeconomic status (Cath et al., 2017). By functioning within the sandbox, developers could optimize their algorithms to eradicate discriminatory practices and improve transparency in decision-making processes (Kera, 2021). This proactive supervision not only averted potential misuse of AI technology but also cultivated public trust in intelligent financial services within the city.

### *5.4 Strengthening Institutional and Legislative Mechanisms for Smart City Governance*

Finally, it is essential to blend institutional mechanisms in the implementation of contemporary smart city technologies. This means that current legislative frameworks for urban design must align with the evolving technical and ethical standards. For example, some Data protection regulations, exemplified by the European Union's General Data Protection Regulation (GDPR), have provided a strong foundation for this field, which requires human intervention in multiple types of automated decision-making (ADM) (Binns & Veale, 2021). In particular, Article 22(1) of the GDPR prohibits decisions that rely exclusively on automated processing. All lawful decisions that have legal or similarly significant impacts on individuals must incorporate human oversight within the data-processing decision-making loop (Lazcoz & De Hert, 2023). This functions as a regulatory measure to address high-risk data processing practices and alleviates issues related to AI transparency and accountability in current ethical challenges.

From the data subject's perspective, residents in smart cities (along with other stakeholders) should have a fundamental right of access to any data collected about them. In this context, the Freedom of Information (FOI) Act is an underutilized tool for enhancing transparency and accountability in smart city governance (Francisca & Araujo, Joaquim Filipe Ferraz Esteves, 2021). FOI laws can help citizens understand which data is being collected and how it is utilized within various smart city applications. By integrating FOI provisions into governance structures, we can address concerns about data usage and foster greater trust between citizens and authorities. In practice, streamlining the FOI request process will be essential going forward. In more detail, Figure 3 illustrates a comparative framework between traditional Freedom of Information (FOI) request process and streamlined version. In the revised model, requests are submitted through adaptable digital portals that automatically produce acknowledgement and tracking information. Advanced AI and machine learning techniques enhance classification by automatically categorizing and prioritizing requests. In addition, this model updated centralized digital repositories and automated redaction tools alleviate administrative burdens related to data retrieval and processing. This decision will be alleviated most of concern and current major challenges for redaction (J. Ma et al., 2022; Wagner & Cuillier, 2023). After then, real-time monitoring of request status, along with proactive email or SMS notifications, enhances engagement and trust among requesters. The new system incorporates feedback mechanisms, including post-request surveys and robust analytics, facilitating iterative improvements in service delivery. We also streamline the compliance and reporting are enhanced through public dashboards that present aggregated performance metrics, thereby improving accountability and transparency. A specialized AI/ML workflow guarantees that algorithms effectively label and assign new cases, while human oversight is essential for verifying the accuracy of these decisions, including redaction result, thus balancing the advantages of automation with ethical and legal protections.

**Figure 3.** Comparative Framework for Traditional vs. updated FOI Processes

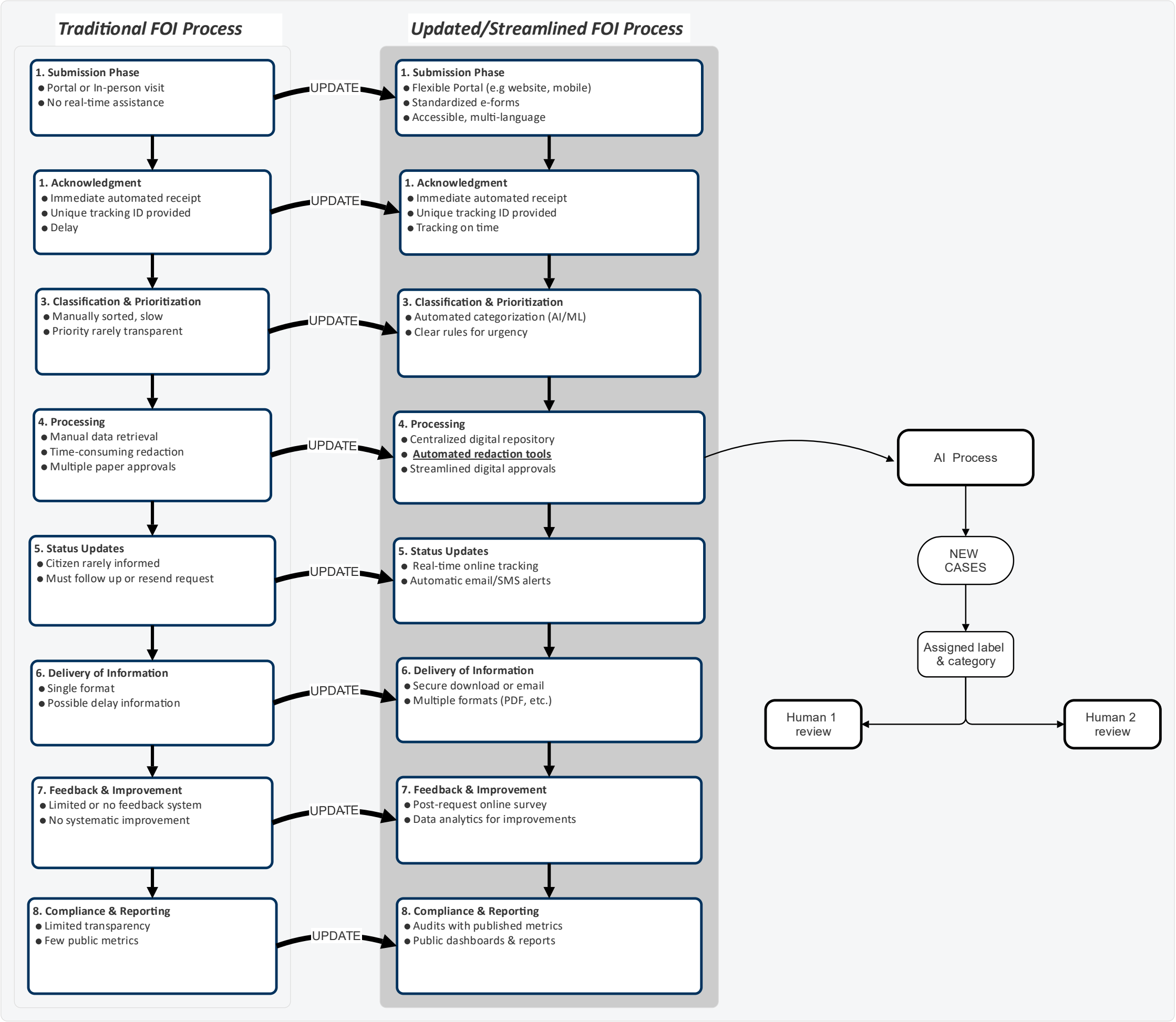

## 6.0 Conclusion

This study has demonstrated that ethical considerations should integral to the governance strategies underpinning smart cities, yet several gaps remain. While the Beard & Longstaff framework has been explored in other domains, this study uniquely applies it to smart city governance. By comparing four principles (e.g., self-determination, fairness, accessibility, purpose) the research uncovers previously overlooked ethical dimensions. Beyond diagnosing ethical pitfalls, the research outlines clear, practice-oriented recommendations for policymakers, city officials, and urban planners. By centering on the Beard & Longstaff framework, it not only proposes methodological improvements but also provides concrete steps (e.g., participatory governance models, regulatory sandboxes) that stakeholders can implement to foster transparent, equitable, and inclusive smart city initiatives. However, future studies should expand upon this framework to account for additional dimensions—such as environmental sustainability, labor dynamics, and the rapid evolution of emerging technologies (e.g., AI and blockchain) that increasingly shape the ethical landscape of modern cities.

To enrich these findings, researchers should conduct broader, longitudinal analyses that span diverse socioeconomic contexts, ensuring that lessons gleaned are inclusive and applicable across varying governance models. Efforts must also intensify around participatory methods, particularly by involving marginalized communities whose perspectives remain underrepresented. In doing so, smart city initiatives can more effectively balance the need for innovation with the imperative for equity, transparency, and inclusivity.Ultimately, the strength of any smart city lies not solely in its technological sophistication but in the ethical frameworks guiding its development. By building on the insights provided by Beard & Longstaff and incorporating a wider set of ethical and governance considerations, policymakers and practitioners alike can craft urban environments that truly serve the common good. Through regulatory sandboxes, stakeholder engagement, and thoughtful adoption of emerging technologies, smart cities can evolve into equitable, sustainable, and human-centric ecosystems that uphold the rights, dignity, and well-being of all residents.

# Appendix

| # | Reference (Year) | SD (Self-Determination) | F (Fairness) | P (Purpose) | A (Accessibility) | Research Method | Brief Explanation |
|---|---|---|---|---|---|---|---|
| 1 | Albino, V., Berardi, U., & Dangelico, R. M. (2015) | ✓ | ✓ | ✓ | ✓ | Literature Review / Conceptual | Defines and categorizes smart cities, dimensions, performance—key purpose of smart city research. |
| 2 | Anthopoulos, L., Janssen, M., & Weerakkody, V. (2023) | ✓ | | | | Editorial / Intro to Special Issue | Focuses on citizen-centricity in smart cities → emphasizes self-determination. |
| 3 | Beard, M., & Longstaff, S. (2018) | ✓ | ✓ | ✓ | ✓ | Conceptual / Policy Principles | Proposes ethical-by-design principles to ensure fair and responsible technology. |
| 4 | Binns, R., & Veale, M. (2021) | ✓ | | | | Legal / Policy Analysis | GDPR Article 22 & multi-stage profiling → underscores data-subject autonomy (self-determination). |
| 5 | Brayne, S. (2021) | ✓ | ✓ | ✓ | | Empirical (Mixed Methods) | Examines predictive policing & surveillance, highlighting fairness and social justice issues. |
| 6 | Button, K., & Taylor, S. (2000) | | | ✓ | | Quantitative (Economic Analysis) | Links air transportation infrastructure to economic development purpose in broader urban contexts. |
| 7 | Caragliu, A., Del Bo, C., & Nijkamp, P. (2011) | | | ✓ | | Quantitative (Statistical / Empirical) | Conceptualizes smart city performance in Europe → clarifies purpose and metrics. |
| 8 | Cath, C., Wachter, S., Mittelstadt, B., Taddeo, M., & Floridi, L. (2017) | ✓ | | ✓ | | Conceptual / Comparative Policy Analysis | Frames AI in context of a Good Society, discussing purpose behind AI regulation. |
| 9 | Cavada, M., Rogers, C., & Hunt, D. (2014) | ✓ | ✓ | ✓ | ✓ | Conceptual / Literature Review | Exposes contradictory definitions & measures of "smart city," focusing on conceptual purpose. |
| 10 | Chang, V. (2021) | | ✓ | | | Conceptual / Framework | Proposes an ethical framework for big data in smart cities → addresses fairness and responsibility. |
| 11 | de Gooyert, V., Rouwette, E., van Kranenburg, H., & Freeman, E. (2017) | ✓ | | | | Conceptual / Stakeholder Theory | Emphasizes participatory decision-making and self-determination via stakeholder theory. |
| 12 | Degrande, T., Van Den Eynde, S., Vannieuwenborg, F., Colle, D., & Verbrugge, S. (2021) | | | ✓ | | Techno-Economic Modeling | Discusses cost-benefit approach for C-ITS roadside units → clarifies purpose & feasibility. |
| 13 | Deloitte. (2022). Smart Health Communities | | | | ✓ | Policy / Industry Report | Shows how "smart health" initiatives expand healthcare accessibility. |
| 14 | eHealth, W. G. O. (2010) | | | | ✓ | Global Survey / Policy Report | Focuses on telemedicine developments → improving health access globally. |
| 15 | Farthing, S., Howell, J., Lecchi, K., Paleologos, Z., Saintilan, P., & Santow, E. (2019) | | ✓ | | | Policy Discussion Paper | Links technology to human rights, highlighting fairness and non-discrimination. |
| 16 | Francisca, T.-R., & Araujo, J. F. F. E. (2021) | ✓ | | | | Quantitative (Municipal Data) | Examines freedom of information in local government → fosters citizen self-determination. |
| 17 | García, I., & Kim, K. (2020) | | | | ✓ | Mixed Methods (Quant + Qual) | Focuses on active commute & health equity among Hispanic students → accessibility. |
| 18 | Göbel, R., & Kitzing, S. (2023) | ✓ | | | | Technical / Data Privacy Models | Defines anonymity properties → ensures privacy & autonomy for users. |
| 19 | Graham, S., & Wood, D. (2003) | | ✓ | | | Theoretical / Critical | Addresses surveillance and resulting inequality, raising fairness concerns. |
| 20 | Gregor, S., & Hevner, A. R. (2013) | | | ✓ | | Methodological / Conceptual | Positions design science research for maximum purpose and impact. |
| 21 | Heilweil, R. (2020) | | ✓ | | | Journalism / Policy Commentary | Discusses facial recognition bans → preventing bias & ensuring fairness. |
| 22 | Herrera, J. L., Chen, H.-Y., Berrocal, J., Murillo, J. M., & Julien, C. (2022) | ✓ | | | | Technical / System Design | Outlines context-aware privacy-preserving access control → upholds user autonomy. |
| 23 | Hollands, R. G. (2008) | | | ✓ | | Critical / Conceptual | Critiques "smart city" branding vs. actual objectives → purpose. |
| 24 | Kamargianni, M., Li, W., Matyas, M., & Schäfer, A. (2016) | | | | ✓ | Literature Review / Case Insights | Evaluates mobility services for urban transport → accessibility improvements. |
| 25 | Kamtam, P. (2023) | | | | ✓ | Multiple Case Studies (Qualitative) | Proposes people-centered inclusive approaches → accessibility and engagement. |
| 26 | Kera, D. R. (2021) | | ✓ | | | Conceptual / Policy Analysis | Explores RegTech sandboxes → ensuring fair oversight of algorithms. |
| 27 | Keshavarzi, G., Yildirim, Y., & Arefi, M. (2021) | | | ✓ | | Systematic Literature Review | Surveys “smart city” research across scales → clarifies purpose and definitions. |
| 28 | Kevin, M., Mark, R., & Bernd, S. (2019) | | ✓ | | | Comparative Case Studies | Examines ethics & human rights in smart info systems → fairness. |
| 29 | Khromova, S., Méndez, G., Eckelman, M., Herreros-Cantis, P., & Langemeyer, J. (2024) | | | | ✓ | Possibly Mixed or Theoretical | Focuses on social-ecological-technological vulnerability. Could relate to fairness (equity) and purpose (risk mgmt). Exact classification depends on full text details. |
| 30 | Kitchin, R. (2013) | | | ✓ | | Conceptual / Critical | Explores real-time city & big data rationale → purpose of data-driven urbanism. |
| 31 | Kitchin, R. (2016) | | ✓ | ✓ | | Conceptual / Ethical Framework | Discusses ethical aspects in smart urban science → emphasizes fairness and justice. |
| 32 | Kolotouchkina, O., Barroso, C. L., & Sánchez, J. L. M. (2022) | | | | ✓ | Empirical (Surveys / Interviews) or Conceptual | Tackles the digital divide for people with disabilities → accessibility dimension. |
| 33 | Kontokosta, C. E., & Hong, B. (2021) | | ✓ | | | Quantitative (Data Analysis of 311) | Shows how socio-spatial bias in 311 data can undermine fairness in resource allocation. |
| 34 | Lazcoz, G., & De Hert, P. (2023) | ✓ | | | | Legal / Policy Analysis | Centers on human agency in GDPR & AI Act → highlights self-determination. |
| 35 | Lifelo, Z., Ding, J., Ning, H., Qurat-Ul-Ain, & Dhelim, S. (2024) | | | ✓ | | Conceptual / Futuristic Analysis | Discusses AI-enabled metaverse for sustainable smart cities → forward-looking purpose. |
| 36 | Lim, C., Kim, K.-J., & Maglio, P. P. (2018) | | | ✓ | | Conceptual / Reference Model | Proposes big-data reference models for smart cities → addresses purpose and structure. |
| 37 | Ma, J., Huang, X., Mu, Y., & Deng, R. H. (2022) | ✓ | | | | Technical / Cryptography & Security | Focuses on accountability & transparency in data redaction → user control (self-determination). |
| 38 | Ma, X., Li, J., Guo, Z., & Wan, Z. (2024) | | | ✓ | | Conceptual / Technical Overview | Addresses big data’s role in monitoring/developing smart cities → purpose. |
| 39 | Maksimovic, M. (2020) | | | ✓ | | Conceptual / IoT Solutions | Shows how IoT-based waste management meets environmental and innovation goals → purpose. |
| 40 | Mark, R., & Anya, G. (2019) | | ✓ | | | Comparative Case Studies | Studies AI & big-data ethics in large European cities → focuses on fairness. |
| 41 | Meijer, A., & Bolívar, M. P. R. (2016) | | | ✓ | | Literature Review / Theoretical | Examines smart urban governance → clarifies purpose and aims of governance structures. |
| 42 | Minyard, K. J. (2015) | | | ✓ | | Policy / Commentary | Identifies public health system change as a smart city opportunity → purpose. |
| 43 | Mora, H., Gilart-Iglesias, V., Pérez-Delhoyo, R., Andújar-Montoya... (2016) | | | | ✓ | Technical / System Development | Describes a cloud-based system to analyze urban accessibility. |
| 44 | Musselwhite, C. (2022) | ✓ | | | | Conceptual / Review of Gehl’s Approach | Highlights Jan Gehl’s human-centered urban design → self-determination and participation. |
| 45 | Nam, T., & Pardo, T. A. (2011a, 2011b) | | | ✓ | | Conceptual / Conference Papers | Introduces triad of technology, people, institutions → purpose in smart city conceptualization. (Merged duplicates.) |
| 46 | NDIA. (2018) | | | | ✓ | Advocacy / Data Mapping | Identifies worst connected US cities → underscores digital access deficits. |
| 47 | Negro, S. O., Alkemade, F., & Hekkert, M. P. (2012) | | | ✓ | | Literature Review / Innovation System Analysis | Explores slow diffusion of renewables → ties to sustainability purpose. |
| 48 | OECD. (2020) | | | ✓ | | Policy / Macro-level Analysis | Examines digital economy trends shaping city strategies → purpose. |
| 49 | OECD. (2021) | | ✓ | | | Policy Framework | OECD AI principles → fair and trustworthy AI. |
| 50 | Park, P. D., Blackman, D. A., & Chong, G. (Eds.). (2018) | | | ✓ | | Edited Volume / Educational | Links environmental education with city development → purpose in sustainability. |
| 51 | Park, S. (2017) | | | | ✓ | Qualitative (Interviews / Case) | Explores rural digital exclusion → highlights accessibility challenges. |
| 52 | Puussaar, A., Johnson, I. G., Montague, K., James, P., & Wright, P. (2018) | ✓ | | | | Action Research / Participatory | Shows how open data fosters civic advocacy → user self-determination. |
| 53 | Raghunatha, A., Thollander, P., & Barthel, S. (2023) | | | ✓ | | Policy / Framework Development | Offers a policy framework for drone transportation → purpose in new mobility. |
| 54 | Randell-Moon, H. E. K., & Hynes, D. (2022) | | | | ✓ | Policy / Critical Analysis | Examines rural/regional smart policy → connectivity/accessibility. |
| 55 | Richardson, D. (2024) | ✓ | | | | Technical / Policy Thesis | Addresses security & rights in smart city infra → underscores self-determination. |
| 56 | Ryerson, M. S., Davidson, J. H., Csere, M. C., Kennedy, E., & Reina, V. J. (2022) | | ✓ | ✓ | ✓ | Quantitative (Accessibility Modeling) | Proposes equity-driven planning typologies (fairness), strategic purpose, and accessibility measures. |
| 57 | Sánchez-Corcuera, R., Nuñez-Marcos, A., Sesma-Solance, J., et al. (2019) | | | ✓ | | Literature Survey / Technical Overview | Surveys technologies, domains, challenges for future smart cities → clarifies purpose. |
| 58 | Sennett, R. (2018) | | ✓ | | | Philosophical / Ethical Treatise | Explores urban ethics & inclusiveness → addresses fairness dimension. |
| 59 | Serrano, P., Gramaglia, M., Mancini, F., Chiaraviglio, L., & Bianchi, G. (2023) | | | | ✓ | Technical / Empirical | Studies Google Loon's balloon-based internet → extending accessibility in remote areas. |
| 60 | Sheehy, B., & Ng, Y.-F. (2024) | | ✓ | | | Legal / Policy Analysis | Examines AI-based govt decision-making → stresses fairness and accountability. |
| 61 | Shelton, T., Zook, M., & Wiig, A. (2015a, 2015b) | | | ✓ | | Critical / Conceptual | Explores the "actually existing smart city" → purpose behind real implementations. (Merged duplicates.) |
| 62 | Swargiary, K., & Roy, K. (2023) | | | | ✓ | Theoretical / Policy Perspective | Connects adult education to inclusive development → accessibility dimension. |
| 63 | Szewczenko, A. (2020) | | | | ✓ | Literature Review | Focuses on improving urban-space accessibility for older adults. |
| 64 | Tan, S., & Taeihagh, A. (2020) | | | ✓ | | Systematic Literature Review | Examines smart city governance in developing countries → purpose and frameworks. |
| 65 | Taylor, R., Wanjiru, H., Johnson, O. W., & Johnson, F. X. (2020) | | | ✓ | | Systems Modeling / Stakeholder Analysis | Investigates sustainable charcoal markets → highlights purpose in socio-environmental transitions. |
| 66 | Townsend, A. M. (2013) | | | ✓ | | Book (Conceptual + Case Illustrations) | Explores motivations & rationale behind the smart-city movement → purpose. |
| 67 | Truby, J. (2020) | | ✓ | | | Policy / Regulatory Analysis | Sandbox 2.0 policy reforms in fintech → ensures fairness and equitable outcomes. |
| 68 | Tseng, P.-E., & Wang, Y.-H. (2021) | | ✓ | | | Ethical Analysis | Compares deontological vs. utilitarian reasoning in outbreak policy → fairness dilemmas. |
| 69 | UN-Habitat. (2021) | ✓ | ✓ | ✓ | ✓ | Global Policy / Report | Outlines future city scenarios from a global perspective → emphasizes urban purpose. |
| 70 | Vanolo, A. (2014) | | ✓ | | | Critical / Theoretical | Argues "smartmentality" is a disciplinary strategy → raises fairness & equity concerns. |
| 71 | Vazquez Brust, A. (2019) | | | ✓ | | Book Review / Conceptual | Reviews The City of Tomorrow, focusing on urban innovation purpose. |
| 72 | Wagner, A. J., & Cuillier, D. (2023) | ✓ | | | | Survey Research / Public Admin | Looks at FOI fees that can hinder citizen self-determination in accessing info. |
| 73 | WEF. (2021) | | ✓ | | | Policy Benchmarks / White Paper | Provides ethical & responsible guidelines for smart-city development → fairness. |
| 74 | Woyke, E. (2019) | | | | ✓ | Journalism / Investigative | Warns "smart cities" can be inaccessible for disabled persons → accessibility. |
| 75 | Zakir, M., Wolbring, G., & Yanushkevich, S. (2024) | | | | ✓ | Quantitative (Probabilistic Modeling) | Identifies barriers to accessibility for persons with disabilities via causal modeling. |